\documentclass[%
 reprint,
superscriptaddress,
 amsmath,amssymb,
 aps,
noeprint,
onecolumn
]{revtex4-2}

\usepackage{amsmath}
\usepackage{bbm}
\usepackage{siunitx}
\usepackage{circuitikz}
\usepackage{extarrows}
\usepackage{bbold}
\usepackage{graphicx}
\usepackage{dcolumn}
\usepackage{bm}
\usepackage[colorlinks=true,linkcolor=blue,citecolor=blue,urlcolor=blue]{hyperref}

\newcommand{\beginsupplement}{%
        \setcounter{table}{0}
        \renewcommand{\thetable}{S\arabic{table}}%
        \setcounter{figure}{0}
        \renewcommand{\thefigure}{S\arabic{figure}}%
        \setcounter{equation}{0}
        \def\theequation{S\arabic{equation}}
     }

\begin{document}


\title{Active topolectrical circuits}

\author{Tejas Kotwal}
\affiliation{Department of Mathematics,
Indian Institute of Technology Bombay,
Mumbai 400076, India}
\affiliation{Department of Mathematics, Massachusetts Institute of Technology,
77 Massachusetts Avenue, Cambridge, MA 02139, U.S.A.}
\affiliation{Division of Applied Mathematics, Brown University, 182 George Street, Providence, RI 02912, U.S.A.}
\author{Fischer Moseley}
\affiliation{Department of Physics, Massachusetts Institute of Technology,
77 Massachusetts Avenue, Cambridge, MA 02139, U.S.A.}
\author{Alexander Stegmaier}
\affiliation{Institut f\"ur Theoretische Physik und Astrophysik, Universit\"at W\"urzburg, D-97074 W\"urzburg, Germany}
\author{Stefan Imhof}
\author{Hauke Brand}
\author{Tobias Kie\ss ling}
\affiliation{Physikalisches Institut der Universit\"at  W\"urzburg, Universit\"at W\"urzburg, D-97074 W\"urzburg, Germany}
\author{Ronny Thomale}
\affiliation{Institut f\"ur Theoretische Physik und Astrophysik, Universit\"at W\"urzburg, D-97074 W\"urzburg, Germany}
\author{Henrik Ronellenfitsch}
\affiliation{Department of Mathematics, Massachusetts Institute of Technology,
77 Massachusetts Avenue, Cambridge, MA 02139, U.S.A.}
\affiliation{%
Physics Department, Williams College, 33 Lab Campus Drive, Williamstown, MA 01267, U.S.A.
}%
\author{J\"orn Dunkel}
\email{dunkel@mit.edu}
\affiliation{Department of Mathematics, Massachusetts Institute of Technology,
77 Massachusetts Avenue, Cambridge, MA 02139, U.S.A.}

\begin{abstract}
\textbf{Abstract.} The transfer of topological concepts from the quantum world to classical mechanical and electronic systems has opened fundamentally new approaches to protected information transmission and wave guidance. A particularly promising technology are recently discovered topolectrical circuits that achieve robust electric signal transduction by mimicking edge currents in quantum Hall systems. In parallel, modern active matter research has shown how autonomous units driven by internal energy reservoirs can spontaneously self-organize into collective coherent dynamics. Here, we unify key ideas from these two previously disparate fields to develop design principles for active topolectrical circuits (ATCs) that can self-excite topologically protected global signal patterns.  Realizing autonomous active units through nonlinear Chua diode circuits, we theoretically predict and experimentally confirm the emergence of self-organized protected edge oscillations in one- and two-dimensional ATCs. The close agreement between theory, simulations and experiments implies that nonlinear ATCs provide a robust and versatile platform  for developing high-dimensional autonomous electrical circuits with topologically protected functionalities.
\end{abstract}

\maketitle

\section*{Introduction}
Information transfer and storage in natural and man-made active systems, from sensory organs~\cite{fitzhugh1955,nagumo1962,izhikevich2007} to the internet, rely on the robust exchange of electrical signals between a large number of autonomous units that balance local energy uptake and dissipation~\cite{delRio_2001,delRio_2003}.
Major advances in the understanding of photonic~\cite{Weimann2017,Khanikaev2013,Lumer2013,Noh2018}, acoustic~\cite{yang2015,Xue2019,Ni2018} and mechanical~\cite{kane2014,huber2016,nash2015,Chen2014} metamaterials have shown that topological protection~\cite{hasan2010,bernevig2006,fu2007,hsieh2008,moore2007,qi2011,Kempkes2019,Peterson2018} enables the stabilization and localization of signal propagation in passive and active~\cite{Souslov2016,Woodhouse2018,shankar2020} dynamical systems that violate time-reversal and/or other symmetries. Recent studies  have  successfully applied these ideas to realize topolectrical circuits~\cite{lee2018} in the passive linear~\cite{imhof2018,PhysRevB.99.161114,ningyuan2015,albert2015,luo2018,PhysRevB.98.201402} and passive nonlinear~\cite{Hadad2018,Ezawa2021} regime. However, despite substantial progress in the development of topological wave guides~\cite{Iwamoto:21}, {lasers~\cite{Harari2018,Bandres2018}}, and transmission lines~\cite{wang2019,sayed2018,jiang2019,iyer2008}, the transfer of these concepts to active~\cite{2017FoWoDu_PRL,2016Woodhouse_PNAS} nonlinear circuits made from autonomously acting units still poses an unsolved challenge. From a broader perspective, harnessing topological protection in nonlinear active circuits not only promises a new generation of autonomous devices, but understanding their design and self-organization principles may also offer insights into information storage and processing mechanisms in living systems, which  integrate cellular activity, electrical signaling and nonlinear feedback to coordinate essential biological functions~\cite{NeuronToBrain,2004ConnorsLong}.

\begin{figure*}
    \centering
    \includegraphics[width=.9\textwidth]{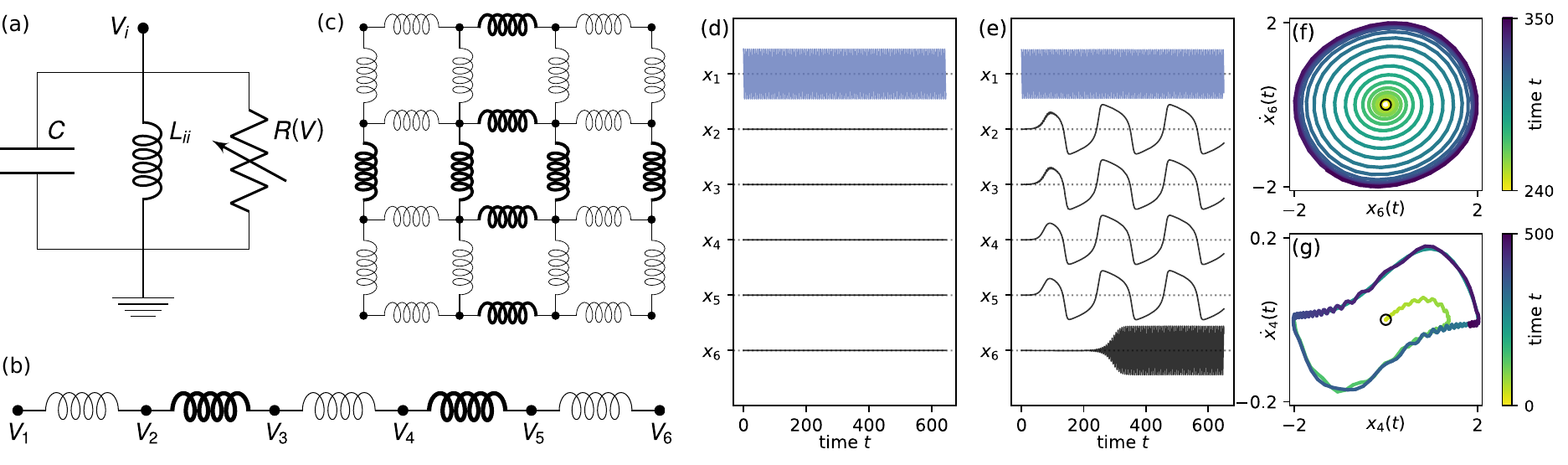}
    \caption{Schematic and dynamics of active topolectrical SSH circuits. (a) Circuit diagram of a basic vdP oscillator with capacitance $C$, inductance $L_{ii}$ and nonlinear resistor $R(V)$.
    (b,c)~1D~and 2D SSH circuits where each node is a vdP oscillator. Thick lines indicate strong coupling $s$ and thin lines weak coupling $w\ll s$.
    (d)~Oscillator dynamics in an undamped passive ($\varepsilon =0$) 1D SSH circuit with 6 nodes. Applying a nonzero initial voltage at the first node, the oscillation remains exponentially localized on that edge.
    (e--g)~Dynamics of a 1D ATC ($\varepsilon >0$) for the same initial condition as in (d).  The high-frequency topological edge mode is activated first and induces slow synchronized bulk oscillations, which eventually actuate the second topological mode at the opposite end. Phase portraits of the boundary (f) and bulk (g) nodes show the limit cycles of the fast and slow oscillations. The edge oscillators
    show approximately circular limit cycles typical of
    weak nonlinearity while the bulk dynamics is of strong
    relaxation type despite $\varepsilon$ being small.
    Simulation parameters in (d-e): $g=1$, $w=0.003, s = g - w$ with  $\varepsilon=0$ in (d) and $\varepsilon=0.1$ in (e-g). Initial conditions: $x_1(0)=2$ and $\dot{x}_1(0)=x_j(0)=\dot{x}_j(0)=0$ for $j\ge 2$.
    \label{fig:1}}
\end{figure*}

\par
 Exploiting a mathematical analogy with active Brownian particle systems~\cite{Woodhouse2018}, we theoretically develop and experimentally demonstrate general design principles for active topolectrical circuits (ATCs) that achieve self-organized, self-sustained, topologically protected electric current patterns. The main building blocks of ATCs  are nonlinear dissipative elements that exhibit an effectively negative resistance  over a certain voltage range. Negative resistances can be realized using van der Pol (vdP) circuits~\cite{2014Krein}, tunnel diodes, unijunction transistors, solid-state thyristors~\cite{gottlieb1997}, or operational amplifiers set as negative impedance converters through current inversion~\cite{hofmann2019}, and the design principles described below are applicable to all these systems. Indeed, we expect them to apply to an even broader class of nonlinear systems, as similar
 dynamics also describe electromagnetic resonators with Kerr-type nonlinearities~\cite{dobrykh2018,Morimoto2016,Shi2017}.

\section*{Results}

\textbf{Theoretical framework.}
Active electronic circuits with basic non-topological interactions have been studied previously as models of neuronal networks~\cite{2014Krein} and solitary signal transport~\cite{delRio_2001,delRio_2003}. To leading order, the negative resistance elements in an active circuit can be described~\cite{delRio_2001,delRio_2003} by a nonlinear Rayleigh-type conductance $R^{-1}(V) = -(\alpha- \gamma V^2)$, where $V$ denotes voltage, and $\alpha$ and $\gamma$ are positive parameters.  A prototypical example is the vdP oscillator circuit with capacitance $C$ and inductance $L_{ii}$, as shown in Fig.~\ref{fig:1}(a). When expressed  in terms of the rescaled dimensionless voltage $\hat{V}_i = x_i = \sqrt{3\gamma/\alpha}\, V_i$, the dynamics of an isolated vdP-unit $i$ is governed by~\cite{stoker1950} (Methods)
\begin{align}
    \ddot x_i - \varepsilon \left(1 - x_i^2\right) \dot x_i+ x_i =0  ,
    \label{eq:vdp-1}
\end{align}
where $\varepsilon = \alpha\sqrt{L/C}$, $t = \sqrt{LC} \hat{t}$, and $\dot{x}_i=dx_i/d \hat{t}$.
{Because this active system
is linearly unstable, a stabilizing nonlinearity is necessary to obtain bounded dynamics. In this case,
we find limit cycle oscillations [Fig.~\ref{fig:1}(f)].}~\eqref{eq:vdp-1} is intimately related to that of a harmonically trapped active Brownian particle~\cite{Romanczuk2012}, with position coordinate $y$ and velocity $u=\dot{y}$ described by the standard cubic friction model $\dot u = \varepsilon (1 - u^2/3) u - y$. Upon taking the time derivative and identifying $u=\dot{y}=x_i$, one recovers \eqref{eq:vdp-1}. It was shown recently~\cite{Woodhouse2018} that suitably coupled mechanical chains of active Brownian particles can autonomously actuate topological modes. This insight provides  guidance for the design of ATCs.

To design an ATC with desired topological properties, we generalize \eqref{eq:vdp-1} by introducing suitably chosen couplings between vdP units $i$ and $j$ through inductances $L_{ij}$; see Fig.~\ref{fig:1}(b,c) for two examples.
Assuming a lattice of vdP units and introducing the symmetric coupling matrix elements $\beta_{ij} = -L_0/L_{ij}$ for $i \neq j$,
and $\beta_{ii} = \sum_{ k } L_0/L_{ik}$,  \eqref{eq:vdp-1} generalizes to
\begin{align}
    \ddot x_i - \varepsilon \left(1 - x_i^2\right) \dot x_i + \sum_j \beta_{ij} x_j = 0,
    \label{eq:vdp}
\end{align}
where $L_0$ is the smallest inductance in the circuit.

\par
Interpreting each individual vdP unit as a node in the network graph [Fig.~\ref{fig:1}(b,c)], the effective dimension~$d$ of the ATC can be tuned by increasing the number of couplings.
In principle, arbitrary values $d\ge 1$ are realizable with large, densely connected networks~\cite{lee2018}.  For example, for $d$-dimensional cubic lattices, rows of the coupling matrix $\boldsymbol{\beta}=(\beta_{ij})$ corresponding to bulk nodes will have $2d$ off-diagonal entries.  Below, we will focus on the cases $d=1$ [Fig.~\ref{fig:1}(b)] and $d=2$ [Fig.~\ref{fig:1}(c)] to demonstrate the implementation and key properties of ATCs. Generally, through an appropriate choice of $\boldsymbol{\beta}$, it is possible to realize a wide range of topological phases.

\begin{figure*}[ht!]
    \centering
    \includegraphics[width=\textwidth]{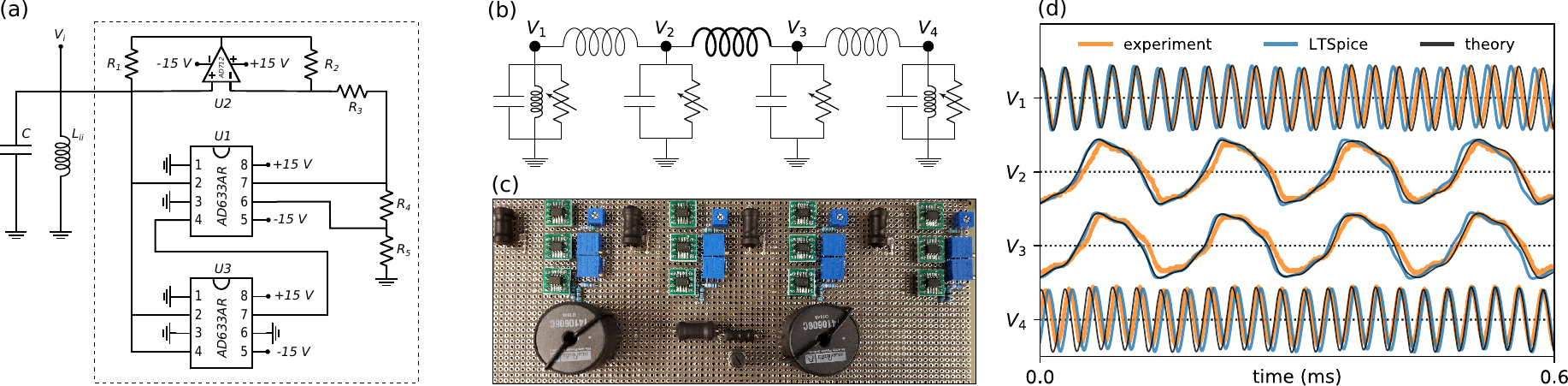}
    \caption{Experimental realization of a 1D active topolectric
    circuit with 4 nodes.
    (a) Implementation of a single active node using
    a Chua diode circuit (Supplementary Information).
    (b) Four-node active topolectrical circuit using $C=42\,\mathrm{nF}$, $L_{12}=L_{34}=10.5\,\mathrm{mH}$, $L_{23}=525\,\mathrm{\mu H}$.
    To obtain the classical SSH coupling matrix Eq. \eqref{eq:beta-6}, no inductors were used
    for the bulk nodes, such that $L_{11} = L_{44}= 525\,\mathrm{\mu H} $.
    The corresponding dimensionless parameters are
    $s=1.0$, $w=0.05$, $g=1.05$, and $\varepsilon =0.28$.
    (c) Experimental setup on circuit board.
    (d) Experimental data (orange lines), simulations of the full circuit
    using LTSpice (blue lines), and theoretical
    simulation using \eqref{eq:vdp} match. Voltages $V_{1}$--$V_4$
    vary between $\pm 6\;\mathrm{V}$, in-phase initial conditions were chosen on the two bulk oscillators.}
    \label{fig:2}
\end{figure*}

\textbf{One-dimensional Su-Schrieffer-Heeger ATC: Theoretical analysis.}
As a first realization of an ATC, we consider the coupling matrix of a one-dimensional (1D) Su-Schrieffer-Heeger (SSH) model. Originally introduced to describe polyacetylene~\cite{su1979}, the SSH model is known to support topologically protected boundary modes in a variety of quantum~\cite{asboth2016} and classical~\cite{Chen2014,Woodhouse2018,lee2018} systems. The $n\times n$ SSH coupling matrix for the case of $n=6$ unit vdPs reads (Methods)
\begin{align}
\label{eq:beta-6}
\boldsymbol{\beta} = \begin{pmatrix} g & -w & 0 & 0 & 0 & 0 \\
 				-w & g & -s & 0 & 0 & 0 \\
				 0 & -s & g & -w & 0 & 0 \\
				 0 & 0 & -w & g & -s & 0 \\
				 0 & 0 & 0 & -s & g & -w  \\
				 0 & 0 & 0 & 0 & -w & g
				 \end{pmatrix},
\end{align}
with the diagonal elements $\beta_{ii}= g>0$ representing the electrical grounds. In electrical circuits, the diagonal elements of the coupling matrix $\beta_{ii} = \sum_k L_0/L_{ik}$ depend
on the neighboring inductors and are generally not uniform.
Therefore, we introduce additional ground inductances
$L_{ii}$ at all boundaries, such that $\beta_{ii} = g$ is
uniform and the coupling matrix represents an instance
of the SSH model. Upon generalizing this coupling matrix by permitting non-uniform grounds~$g_i$, one  can achieve additional frequency control~(Supplementary Information).
The off-diagonal entries  $w>0$ and $s>w>0$ encode the weak and strong couplings, respectively.
Adding more units extends the matrix symmetrically along the diagonal.  As in the original quantum SSH model, the strongly coupled nodes form `dimers' which are connected to  each other through weak bonds [thin lines in Fig.~\ref{fig:1}(b)].

To gain intuition about the dynamics of the 1D SSH circuit, it is instructive to first consider the undamped passive topolectrical SSH circuit with $\varepsilon=0$ and $s\gg w$. In this case,~\eqref{eq:vdp} and~\eqref{eq:beta-6} reduce to a linear dynamical system that exhibits topologically protected, exponentially localized modes at both ends of the chain~\cite{asboth2016}. These edge modes have finite-frequency for $g>0$ and become zero-modes as $g\to 0$. Their presence can be illustrated by considering an ideal passive ($\varepsilon=0$) circuit~\cite{lee2018} in the almost fully dimerized topological regime ($w\ll s$) with all nodes initially at rest, $x_i(t)= \dot x_i(t) =0$ for $t<0$. Upon initializing the chain by imposing a nonzero voltage value at the left edge, $x_1(0) >0$, the boundary node will oscillate with a non-zero amplitude at frequency $\sqrt{g}$, while the amplitudes of the other nodes remain exponentially small~[Fig.~\ref{fig:1}(d)].  In the next part, we will see that ATCs with $\varepsilon>0$ exhibit fundamentally different behaviors that promise a broad range of novel applications.

Adding just a small amount of activity ($0<\varepsilon \ll \sqrt{g} $) significantly alters the dynamics of the edge and bulk oscillators [Fig.~\ref{fig:1}(e--g)]. In the topological regime, characterized by $s \gg w $ and  $g \ge s + w$, the boundary nodes are only weakly coupled to the bulk and behave similarly to isolated vdP oscillators. The active local energy input renders the rest state ($x_i =\dot x_i = 0$) unstable, forcing the boundary nodes to approach a stable limit cycle corresponding to an oscillation at frequency~$\sqrt{g}$~[Fig.~\ref{fig:1}(e,f)]. By contrast, bulk nodes are strongly coupled to one or more neighbors, resulting in a non-topological, distinct bulk dynamics
reminiscent of highly nonlinear relaxation oscillations~[Fig.~\ref{fig:1}(e,g)].
For fully decoupled pairs of vdP oscillators, one expects
stable phase-locked solutions with relative phase~$0$ or~$\pi$~\cite{Storti1982,wirkus2002}.
{Because the system is close to the vdP oscillator's
Andronov-Hopf bifurcation~\cite{strogatz2018}, the frequencies of the limit
cycle are approximately given by the imaginary
parts of the Jacobian of the dynamics \eqref{eq:vdp} at
the unstable rest state.
The Jacobian's eigenvalues are} $\lambda_{k,\pm} = \frac{1}{2}(\varepsilon \pm \sqrt{\varepsilon^2 - 4\mu_k}) =\pm \sqrt{-\mu_k} + \mathcal{O}(\varepsilon)$, where $\mu_k$ are the eigenvalues of the coupling matrix $\boldsymbol{\beta}$ (Supplementary Information). {This way, topological
modes encoded in $\boldsymbol{\beta}$ can
appear in the nonlinear regime.} In the fully dimerized limit ($w\rightarrow 0$), $\mu_k$ can be calculated explicitly and one finds two degenerate eigenstates at $\mu_k=g$, corresponding to the topologically protected edge modes, and $n-2$ degenerate bulk modes at $\mu_k=g\pm s$ (Supplementary Information). The bulk eigenvectors are pairwise localized  with components $(1, \pm 1)$ on one of the dimers, and zero everywhere else, corresponding to a low-energy in-phase oscillation ($\mu_k=g-s$) and a high-energy anti-phase ($\mu_k=g+s$) oscillation (Supplementary Information). This implies that the dimers' dynamics near the rest state is approximately decoupled and governed by the eigenvalues $\sqrt{-g\mp s}$. If $g>s$, all modes are oscillatory, while for $g<s$, the anti-phase mode becomes unstable and is not physically realizable.
Armed with these analytical insights, we proceed to numerically characterize the interesting nonlinear effects that arise in ATCs. In the simulations, we can fix $g=1$ without loss of generality, which is equivalent to dividing \eqref{eq:vdp} by $g$ and rescaling~$t \to t/\sqrt{g}$, $\varepsilon \to \varepsilon/\sqrt{g}$, $s \to s/g$ and~$w \to w/g$.

In contrast to passive ($\varepsilon=0$) topolectrical circuits, which remain quiescent in the bulk when initiated at the edge [Fig.~\ref{fig:1}(d)], ATCs with $\varepsilon>0$ exhibit complex self-organization and synchronization phenomena. To explain the underlying physical mechanisms, we consider the 1D SSH ATC from Fig.~\ref{fig:1}(b) with the same initial condition as for the passive circuit in Fig.~\ref{fig:1}(d). When one edge node of the ATC is initialized at time $t=0$, it rapidly settles into a limit-cycle oscillation~[Fig.~\ref{fig:1}(e,f)], as predicted above. As it gets activated, the edge mode imparts a weak external forcing on the neighboring
strongly-coupled dimer. The combination of forcing and local energy input from the negative resistance drives the dimer into its low-energy state, characterized by a slow nonlinear in-phase oscillation of the dimer nodes with period  $2\pi/\sqrt{g-s}$~[Fig.~\ref{fig:1}(e,g)]. The first bulk dimer then in turn activates the second dimer, and so on, resulting in a globally synchronized bulk state. The activation front eventually reaches the last node, where it triggers the second topological edge mode [Fig.~\ref{fig:1}(f)]. The final state of the chain is a robust attractor that is  self-sustained and could be used for active solitary signal transmission.

\begin{figure}
    \centering
    \includegraphics[width=0.5\columnwidth]{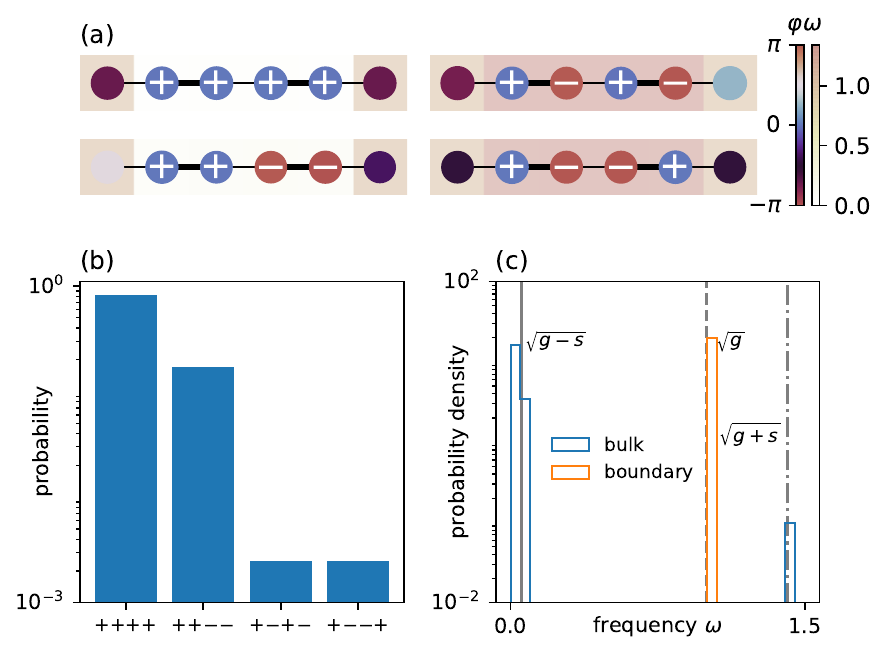}
    \caption{Attractor statistics of a 1D ATC. (a) An active SSH circuit with $n=6$ vdP nodes has four qualitatively distinct stable attractors. Circle size represents instantaneous absolute voltage $|x_i(t)|$, symbol color the phase $\varphi$ and background color the time-averaged frequency $\omega = \langle\dot\varphi\rangle$ of the oscillators. Two low-energy attractors exhibit in-phase dynamics of the bulk dimer nodes,
    and two high-energy attractors show anti-phase dynamics on the bulk dimers.
    (b)~Low-energy attractors are substantially more frequently realized when bulk nodes are initialized with uniformly random $x_i(0)\in [-4,4]$ and $\dot x_i(0)\in[-0.4,0.4]$. The two in-phase states
    are approached for $\approx 99.5\%$ of initial conditions.
    (c)~Frequencies of nonlinear bulk oscillations cluster around  $\sqrt{g\pm s}$ (solid/dash-dotted line), boundary oscillations around $\sqrt{g}$ (dashed line), as predicted by linearized theory. Simulation parameters: $g=1, w=0.003, s=g-w$, and $\varepsilon=0.1$. Histograms were computed from $50,000$ trajectories, integrated up to $t=1000$.
    \label{fig:3}}
\end{figure}

\par
\textbf{1D Su-Schrieffer-Heeger ATC: Experimental realization.}
As a first experimental demonstration of the underlying general concepts, we built a 1D 4-node ATC
using a Chua diode circuit as the active element~[Fig.~\ref{fig:2}(a,b)]. The measured time-series of the oscillator voltages exhibit the theoretically predicted topological edge modes and low-frequency bulk dynamics  (Fig.~\ref{fig:1}). In the case of active circuits, the edge modes excite the bulk abruptly over the scale of just a single
oscillator, even when corresponding
passive circuits would exhibit exponential decay over a longer
scale. The quantitative agreement of the experiments with theory as well as explicit circuit simulations~(Supplementary Information) ~confirms that \eqref{eq:vdp} provides a predictive framework for ATCs -- and that it is straightforward to realize topolectrical materials with off-the-shelf components.
Indeed, the example in Fig.~\ref{fig:2} is only one of many possible ATC implementations~(Supplementary Information).

\begin{figure*}[t]
    \centering
    \includegraphics[width=.95\textwidth]{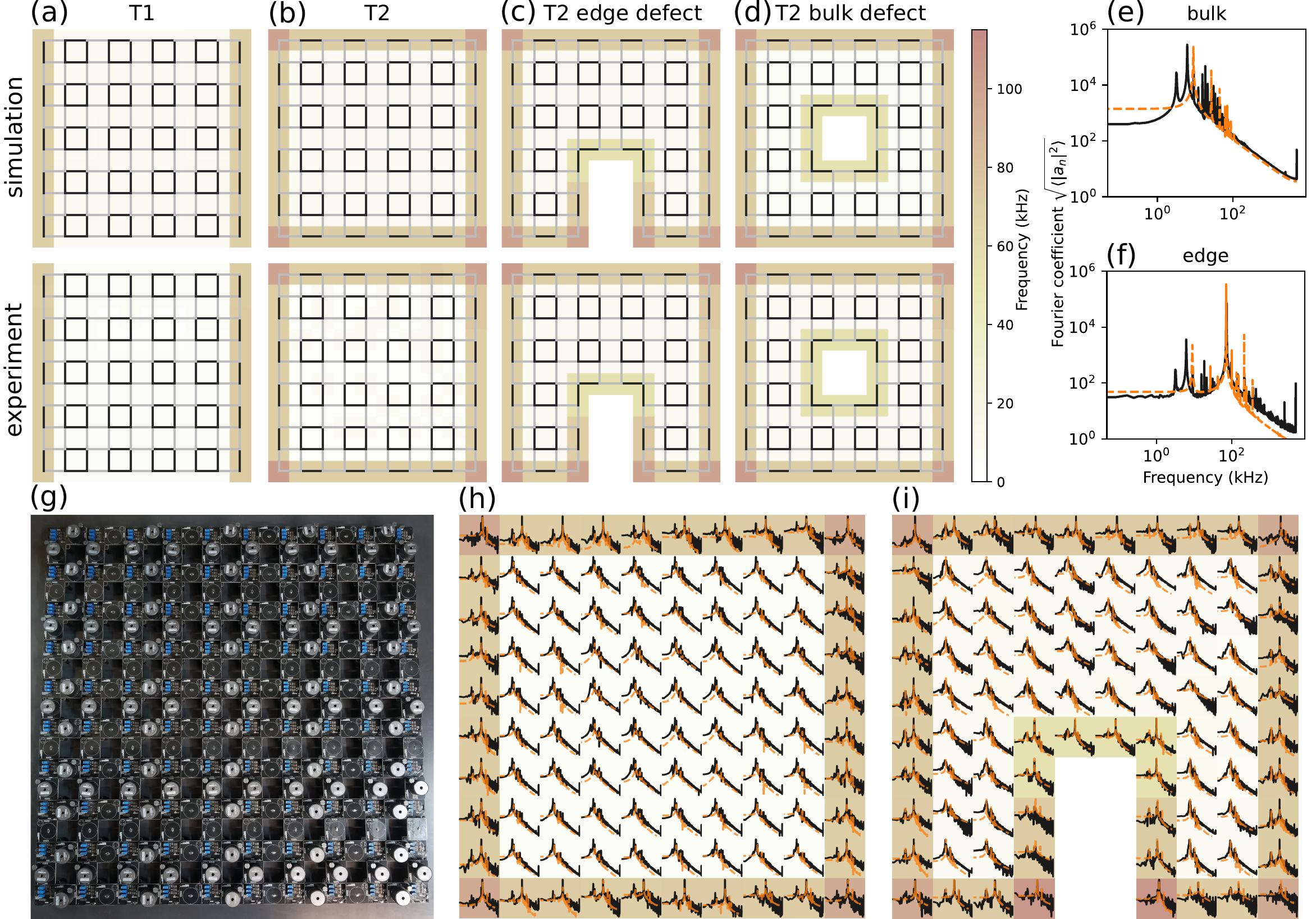}
    \caption{Self-organized self-sustained nonlinear oscillations in 2D ATCs recapitulate topological edge mode phenomena in simulations of \eqref{eq:vdp} and corresponding experiments based on Chua-diode circuits.
    (a,b)~Frequency patterns measured in simulations and experiments confirm the self-excitation of the topological edge mode T1-regime (a) and the topological corner-edge mode T2-regime (b), respectively. Line width indicates inductive coupling strength, background color shows dominant frequency
    obtained from Fourier spectra of the individual oscillator time series.
    (c,d)~In the presence of edge defects (c) or bulk defects (d), the high-frequency dynamics remains robustly localized along the boundaries.
    (e,f) Root-mean-square Fourier spectra measured in experiments
    (black) and simulations (orange) obtained by averaging over all bulk and edge oscillators, respectively, for the T2-networks (b,h). The peak frequencies indicate slow bulk node oscillations and fast edge node oscillations.
    (g)~Experimental ATC setup from which the data shown in  (a--d,h,i) were acquired. An alternative experimental 2D ATC realization is given in the Supplementary Information.
    (h,i) Fourier power spectra for the individual oscillator voltage time-series measured in experiments (b,c bottom; black) and corresponding simulations (b,c top; orange) of \eqref{eq:vdp}. Backgrounds colors show peak frequencies using same color scale as in (a--d). Axis scales of the individual spectra are equal to those in (e,f). Experiments and simulations were performed for grounding capacitances $C=42\,\mathrm{nF}$, bulk inductances $L_w = 13.3\,\mathrm{mH}$ and $L_s = 122\,\mathrm{\mu H}$, and
    nonlinear resistance parameters $\alpha=2.5\,\mathrm{m\Omega}$,
    $\gamma = 90.7\,\mathrm{\mu\Omega\; V^{-2}}$. Despite its minimal generic character, \eqref{eq:vdp} correctly predicts the experimentally observed edge mode phenomenology.
  }
    \label{fig:4}
\end{figure*}

\textbf{Attractor statistics.}
In practice, ATCs can be actuated with a wide range of initial conditions that can select different types of stable asymptotic behaviors (Fig.~\ref{fig:3}), similar to information storage in Hopfield networks. To investigate the likelihood and characteristics of possible attractors, we used \eqref{eq:vdp} to simulate an active SSH circuit with $n=6$ nodes with non-zero initial conditions on the bulk. Since the edge node initial conditions become negligible in this case due to the weak edge-bulk coupling, we fixed zero-initial conditions at the edges. Thus, the topological edge modes are actuated by the bulk dynamics in these simulations. Examples of low-energy and high-energy attractors with slow and fast bulk dimer oscillations, obtained with binary initial conditions in the bulk, $x_i(0) = \pm 0.1$ and $\dot{x}_i(0) =0$, are shown [Fig.~\ref{fig:3}(a)].
These attractors can be classified in terms of the relative signs of the $n_b=n-2=4$ bulk voltages $x_i$ at a fixed time.
Normalizing by the sign of the first bulk oscillator,
we numerically find that there exist four different stable
attractors corresponding to different possible combinations of in-phase and anti-phase dimer oscillations [Fig.~\ref{fig:3}(a)].
\par
To estimate the likelihood of observing a specific attractor in experiments, we performed 50,000 simulations with randomly chosen bulk initial data. To explore a large neighborhood of the bulk limit cycle~[Fig.~\ref{fig:1}(g)], initial conditions were sampled uniformly from the 8-dimensional domain $x_i(0)\in [-4, 4]$, $\dot x_i(0)\in [-0.4,0.4]$. These simulations predict that, in practice, low-energy states with in-phase bulk dimers are much more likely than high-energy attractors with many anti-phase dimers~[Fig.~\ref{fig:3}(b)]. Mixed in-phase/anti-phase dimer attractors are not observed for this parameter regime.
Furthermore, even though the shape of the bulk limit cycles indicates a highly nonlinear relaxation dynamics~[Fig.~\ref{fig:1}(g)], the numerically determined time-averaged frequencies $\langle \dot\varphi\rangle$, where $\varphi(t) = \arctan(\dot x(t)/x(t))$, agree remarkably well with the bulk oscillation frequencies $\sqrt{g\pm s}$ and boundary oscillation frequency $\sqrt{g}$ predicted by the linearized theory above~[Fig.~\ref{fig:3}(c)]. Based on these observations, one expects that low-energy attractors with in-phase bulk dimer dynamics will also be dominant in more complex ATCs, and that their bulk and edge frequencies can be estimated from spectra of the coupling matrix $\boldsymbol{\beta}$ in the weak coupling limit.

\par
\textbf{Experimental realization of a 2D SSH ATC.}
ATC implementations become particularly powerful in $d\ge 2$ dimensions.
Different types of 2D SSH lattices can be constructed by stacking 1D SSH chains and connecting them using alternating strong and weak couplings (Fig.~\ref{fig:4}).
This procedure allows the realization of two essentially different topological
regimes T1 and T2, which
we realized in simulations and experiments~(Supplementary Information):
In the T1-regime, topologically protected modes exist along
two opposite edges of a finite sample [Fig.~\ref{fig:4}(a); SI Movie~1], whereas in the T2-regime
such modes exist on all four of the edges including the corners~[Fig.~\ref{fig:4}(b), SI Movie~2]~\cite{imhof2018}.
Because essential aspects of the dynamics are qualitatively similar for both cases, we focus our discussion on the T2-regime. In this case, one can distinguish three different groups of oscillators: corner
oscillators that are only weakly coupled to their neighbors, strongly coupled vdP dimers along the edges that are similar to the 1D case, and quartets of strongly
coupled vdP oscillators in the bulk [Fig.~\ref{fig:4}(b,e,f,h)]. Each quartet is weakly coupled to its neighbors.
Analogous to the 1D case, the linear stability
of the quartets can be analyzed independently in the weak-coupling limit $w\to 0$, revealing that the lowest-frequency mode is an in-phase state $(1,1; 1,1)$ with period $2\pi/\sqrt{g - 2s}$ (Supplementary Information).
Similar to the 1D case, in the limit $(g - 2s) \to 0^+$, the bulk quartets collectively synchronize to the low-frequency in-phase state, avoiding mixing of bulk excitations
with corners and edges~\cite{Benalcazar2020}.
However, the dimers on the boundary now oscillate at a frequency lower than that of the corner nodes, because  $(g - s) \to s$ as $(g - 2s) \to 0^+$. This opens the intriguing possibility of using topological protected modes to  control oscillation patterns in 2D ATCs. In particular, by varying the ground inductances of each of the nodes one can control frequencies of each of the corner nodes, edge dimers, and bulk quartets (Supplementary Information).

\par
Initializing the 2D active SSH circuits with a nonzero voltage at one of the corner nodes, one finds that essential qualitative features of the
dynamics seen in 1D ATCs carry over to the
2D case. In particular, the boundary nodes belonging to topologically protected corner modes and edge modes become activated one after the other and settle down in their respective vdP limit cycles. Similarly, in the bulk, quartets of strongly coupled oscillators synchronize. Because the in-phase state is the lowest-energy attractor above the quiescent state, the bulk synchronizes in a global in-phase pattern~[Fig.~\ref{fig:4}(b-d)]. Thus, in both 1D and 2D ATCs, topologically protected edge modes become activated via self-sustained oscillations,
while the bulk dynamics is almost decoupled, leading to synchronization.

\par
\textbf{Robustness of self-excited edge modes.}
Crucially, this ATC self-organization principle remains valid in the presence of lattice defects,
demonstrating that topological protection phenomena can survive in the nonlinear regime $\varepsilon>0$~[Fig.~\ref{fig:4}(c,d,i)]. Introducing an edge defect in a passive $(\varepsilon=0)$ 2D SSH grid by removing a few unit cells does not affect the localized nature of the edge state, which now wraps around the defect due to topological protection from the linear coupling~[Fig.~\ref{fig:4}(c,i)]. Self-sustained oscillations in nonlinear active $(\varepsilon>0)$ circuits inherit  this topological protection globally: in the presence of edge defects, all boundary nodes continue to oscillate at a high non-zero frequency while bulk quartets synchronize at low frequency~[Fig.~\ref{fig:4}(b,e),
SI Movie~3]. Similarly, bulk defects also lead to localized nonlinear edge
oscillations~[Fig.~\ref{fig:4}(d), SI Movie~4]. These results show that the SSH network topology can be used to precisely control the individual and collective behavior of coupled nonlinear oscillators.  Furthermore, the above ideas can be extended to achieve control of active traveling wave patterns by means of non-topological defects. By strategically placing bulk defects, one can guide the self-organization of active wave patterns that can be initiated from a single corner node~[Supplementary Information, SI Movie~5]. More broadly, these results open a path towards the  inverse design of functionalized active topolectrical networks~\cite{Ronellenfitsch2020} in the future.

\par

\textbf{General design rules for ATCs.}
{The above analysis, combined with insights from earlier studies~\cite{delRio_2001,delRio_2003,2017FoWoDu_PRL,Woodhouse2018}, suggests general design principles for ATCs, by additively combining local Rayleigh-type activity with suitably designed conservative node coupling interactions ~\cite{Ronellenfitsch2018,Ronellenfitsch2020}. Previous investigations~\cite{2001DuEbErMa,delRio_2003,2017FoWoDu_PRL,Ronellenfitsch2020} showed that, in the low-to-moderate activity regime, nonlinear  Rayleigh-type driving mechanisms often select dynamical attractors that reflect the mode structure of the corresponding non-driven system. In these cases, the nonlinear driving determines the mode amplitudes while quadratic expansions of the interaction potentials determine the mode characteristics. Conversely, recent studies~\cite{Woodhouse2018,Ronellenfitsch2020} have demonstrated that, by designing  the linear response of nonlinear mechanical   networks, one can control the spectral features, such as band gaps, of self-excited waves in the presence of activity. Building on this general idea, the above experimental realizations of ATCs provide an extension to lattices with structured linear couplings that support self-sustained topologically protected phononic excitations. An interesting direction for future research is the design and implementation of autonomous electrical circuits that combine structured nonlinear (e.g. Toda-like~\cite{delRio_2001,Ezawa2021}) couplings with nonlinear  activity, which can be expected to support self-sustained solitonic~\cite{delRio_2001,2017FoWoDu_PRL} edge excitations.
}

\section*{Conclusion}

Active topolectrical circuits promise a wide range of applications, from active wave guides to autonomous electronic circuits with topologically protected functionalities. The framework developed here can be integrated with recently developed methods for the inverse design of network-based metamaterial structures~\cite{2018FoEtAl,Ronellenfitsch2018}, to optimize and tailor the node couplings and transmission properties. The close agreement with our 1D and 2D experiments suggests that the generic ATC model from \eqref{eq:vdp} provides a useful theoretical basis for the implementation of more complex ATCs. For example, by designing the coupling matrix $\boldsymbol{\beta}$ such that the associated dynamical matrix possesses chiral~\cite{hofmann2019} or other symmetries, one can realize different topological phases by utilizing  generic vdP-type nonlinearities.
Another intriguing prospect is the possibility of creating and studying electronic metamaterials with effective dimensions $d>3$ that appear to be difficult to access otherwise. Beyond man-made devices, the above results suggest that it would be interesting to explore whether active biological circuits~\cite{Murugan2017,Knebel2020} may use topological coupling regimes to facilitate robust signal transport and information storage.

\section*{Materials and Methods}

\subsection*{Non-dimensionalization of variables and parameters}
Given a lattice circuit, we state the equations of its voltage dynamics using Kirchoff's laws as follows:
\begin{equation}
	C \ddot{V}_{i} - (\alpha - 3 \gamma V_{i}^{2}) \dot{V}_i + \bigg( \frac{1}{L_{ii}} + \sum_{ \substack{ k \\ k \neq i } } \frac{1}{L_{ik}} \bigg) V_{i} - \sum_{ \substack{ j \\ j \neq i } } \frac{V_j}{L_{ij}} = 0,
\end{equation}
where $V_i$ is the voltage, $C$ is the capacitance, $L_{ii}$ is the ground inductance, and $\alpha$ and $\gamma$ are the van-der-Pol (vdP) parameters of the oscillator at node $i$, while $L_{ij}$ are the coupling inductances between nodes $i$ and $j$. The coupling inductances can be either of $L_w$ or $L_s$ where the subscripts $w$ and $s$ stand for weak and strong coupling respectively ($1/L_{w} < 1/L_{s}$). Note that for nodes $i$ and $k$ that are not connected, we have $1/L_{ik} = 0$.

We introduce a voltage scale $V_i = V_0 \hat{V}_i$, a time scale $t = \tau \hat{t}$, and we scale all the inductances by the smallest inductance, say $L_{0}$, to get $L_{ij} = L_{0} \hat{L}_{ij}$ for all $i$ and $j$. For $\tau$, we use the natural time scale of an oscillator such that $\tau = \sqrt{L_{0}C}$, and the voltage scale is given by $V_0 = \sqrt{\alpha/(3 \gamma)}$. The dimensionless dynamics is then given by \eqref{eq:vdp} where $\hat{V}_i = x_i$, $\varepsilon = \alpha \sqrt{L_0/C}$, $\beta_{ij} = -L_0/L_{ij}$ for $i \neq j$, and $\beta_{ii} = L_0(1/L_{ii} + \sum_{ \substack{ k \\ k \neq i } } 1/L_{ik})$.

In ~\eqref{eq:beta-6}, the weak and strong couplings are given by $w = L_0/L_w$ and $s = L_0/L_s$ respectively, while $L_{ii}$ are chosen such that $g =  L_0(1/L_{ii} + \sum_{ \substack{ k \\ k \neq i } } 1/L_{ik})$ for each $i$. The coupling matrix can be generalized by replacing $g$ by $g_i$.

\subsection*{Circuit implementation}
A detailed description of the two circuit designs based on Chua's diode~\cite{zhong1994} and nonlinear resistors~\cite{delRio_2001}, respectively,  and the component parameters used in our experiments is given in the Supplementary Information.

\acknowledgments{J.D. would like to thank the Isaac Newton Institute for Mathematical Sciences for support and hospitality during the program \lq The Mathematical Design of New Materials\rq ~(supported by EPSRC grant EP/R014604/1) when work on this paper was undertaken. This work was supported by a James S. McDonnell Foundation Complex Systems Scholar Award (J.D.), the MIT Solomon Buchsbaum Research Fund (J.D.) and Robert E. Collins Distinguished Scholar Fund (J.D.). The work in W\"urzburg is funded by the Deutsche Forschungsgemeinschaft (DFG, German Research Foundation) through Project-ID 258499086 - SFB 1170 and through the W\"urzburg-Dresden Cluster of Excellence on Complexity and Topology in Quantum Matter -- \textit{ct.qmat} Project-ID 39085490 - EXC 2147.

J.D. conceived the project. T.Ko. and H.R. performed numerical simulations and analytical calculations and analyzed data. F.M., A.S., S.I, H.B, T.Ki. constructed and analyzed experiments. All authors contributed to writing the manuscript.
T.Ko. and F.M. contributed equally and are joint first authors.
}

\bibliography{active_topolectrics.bib}

\newpage
\beginsupplement

\begin{center}
    \textsc{\textbf{Supplemental Information}}
\end{center}

\section{Eigenvalues of the coupling matrix}
We analyze the linear stability of the quiescent state ($x_i = 0$) of Eq.~(2) in the main text. We find for small deviations, by setting $\dot{x} = y$
\begin{equation}
	\begin{pmatrix} \dot{x} \\
				 \dot{y} \\ \end{pmatrix}
	=
	\begin{pmatrix} 0 & \mathbbm{1} \\
				-\beta & \varepsilon \mathbbm{1} \\ \end{pmatrix}
	\begin{pmatrix} {x} \\
				{y} \\ \end{pmatrix}.
\end{equation}
The eigenvalue equation immediately gives us
\[ \lambda^2 - \varepsilon \lambda = - \mu, \]
where $\mu$ is an eigenvalue of $\beta$ and $\lambda$ is an eigenvalue of the Jacobian. Thus, we obtain
\[ \lambda = \frac{1}{2} \bigg( \varepsilon \pm \sqrt{\varepsilon^2 - 4 \mu} \bigg), \]
to conclude that the most unstable mode occurs for $\mu$ minimal.

First, we compute the eigenvalues of the coupling matrix.
We define
\begin{align}
  B_n = \left(
\begin{array}{cccccc}
 g & -w & 0 & \dots &  & 0 \\
 -w & g & -s &  &  &  \\
 \vdots  & -s & \ddots &  &  &  \\
  &  & -w & g & -s &  \\
  &  &  & -s & g & -w \\
0  & \dots  &  &  & -w & g \\
\end{array}
\right),
\qquad\qquad
C_{n-1} = \left(
\begin{array}{ccccc}
 g & -s & \dots &  & 0  \\
 -s & \ddots &  &  &  \\
 \vdots & -w & g & -s &  \\
  &  & -s & g & -w \\
 0  & \dots &  & -w & g \\
\end{array}
\right),
\end{align}
where $B_n$ is an $n\times n$ matrix and $C_{n-1}$ is an $(n-1)\times (n-1)$ matrix.
Then the characteristic polynomial satisfies the recursion,
\begin{subequations}
\begin{align}
  \chi(B_n,\mu) &= (g -\mu) \chi(C_{n-1},\mu) - w^2 \chi(B_{n-2},\mu), \\
  \chi(C_{n-1},\mu) &= (g - \mu) \chi(B_{n-2},\mu) - s^2 \chi(C_{n-3}, \mu),
\end{align}
\end{subequations}
with initial conditions $\chi(B_2, \mu) = (g - \mu)^2 - w^2$ and
$\chi(C_1,\mu) = g - \mu$.
For simplicity we solve them in the fully dimerized limit, i.e., in the
limiting case, $w=0$, we find that
\begin{align}
  \chi(B_n,\mu) = (g - \mu)^2((g - \mu)^2 - s^2)^{n/2 - 1},
\end{align}
with two modes corresponding to eigenvalue $g$ (the topological edge modes), and rest of the modes corresponding to the symmetric eigenvalues $g \pm s$.

The eigenvectors $x_i$ belonging to the eigenvalue $g$ are $\begin{pmatrix} 1 & 0 & \dots & 0 \\ \end{pmatrix}^\intercal$ and $\begin{pmatrix} 0 & \dots & 0 & 1 \\ \end{pmatrix}^\intercal$ which are precisely the topological edge modes. Further, the eigenvectors corresponding to $g - s$ are the modes
\begin{align*}
\begin{pmatrix} 0 & 1 & 1 & 0 & \dots & 0 \\ \end{pmatrix}^\intercal, \qquad
\begin{pmatrix} 0 & 0 & 0 & 1 & 1 & 0 & \dots & 0 \\ \end{pmatrix}^\intercal, \qquad
\ldots \qquad
\begin{pmatrix} 0 & \dots & 0 & 1 & 1 & 0 \\ \end{pmatrix}^\intercal,
\end{align*}
and those belonging to the eigenvalue $g + s$ are the modes
\begin{align*}
\begin{pmatrix} 0 & 1 & -1 & 0 & \dots & 0 \\ \end{pmatrix}^\intercal, \qquad
\begin{pmatrix} 0 & 0 & 0 & 1 & -1 & 0 & \dots & 0 \\ \end{pmatrix}^\intercal, \qquad
\ldots \qquad
\begin{pmatrix} 0 & \dots & 0 & 1 & -1 & 0 \\ \end{pmatrix}^\intercal.
\end{align*}

The most unstable modes are those with the minimum eigenvalue, which are the degenerate ones belonging to $g - s$. Thus, this explains why the in-phase state is the probabilistically more preferred state. This concludes the analysis near the quiescent state.

We have shown that the quiescent state is unstable by finding the `most preferred' eigenvectors if perturbed. We now analyze the behavior of the system away from the quiescent state. To simplify analysis, it is often useful to consider the system in the fully dimerized limit ($w = 0$). This amounts to analyzing sub-circuits made up of nodes that are only strongly coupled to each other. In essence, we are treating the `strong' sub-circuit as independent to the rest of the network since it is connected to the rest of the network only through weak bonds (in other words, assume the weak bonds have zero strength).

\section{Fully dimerized limit}

For the 1D SSH chain, in the fully dimerized limit, the system decouples into two isolated oscillators from the boundary nodes, and the rest--oscillator pairs from the bulk, called dimers. The dynamics of an isolated oscillator is given by
\begin{equation}
	\ddot{x} - \varepsilon (1 - x^2) \dot{x} + g x = 0,
\end{equation}
whose natural frequency is $\sqrt{g}/(2 \pi)$. In the topological state, the edge nodes of the SSH chain will oscillate with this frequency. The internal dynamics of a dimer however are given by the equations
\begin{subequations}
\begin{align}
	\ddot{x}_1 - \varepsilon (1 - x_1^2) \dot{x}_1 + gx_1 = s x_2, \\
	\ddot{x}_2 - \varepsilon (1 - x_2^2) \dot{x}_2 + gx_2 = s x_1,
\end{align}
\end{subequations}
where $x_1$ and $x_2$ are the voltages of two adjacent nodes that are connected through a strong bond. It can be shown using eigenmode analysis, that the eigen-directions of the coupling matrix are $\begin{pmatrix} 1 & 1 \\ \end{pmatrix}^\intercal$ (in-phase) and $\begin{pmatrix} 1 & -1 \\ \end{pmatrix}^\intercal$ (anti-phase) with frequencies $\sqrt{g - s}/(2 \pi)$ and $\sqrt{g+s}/(2 \pi)$ respectively.

\begin{figure}[t]
    \centering
    \includegraphics[width=.5\columnwidth]{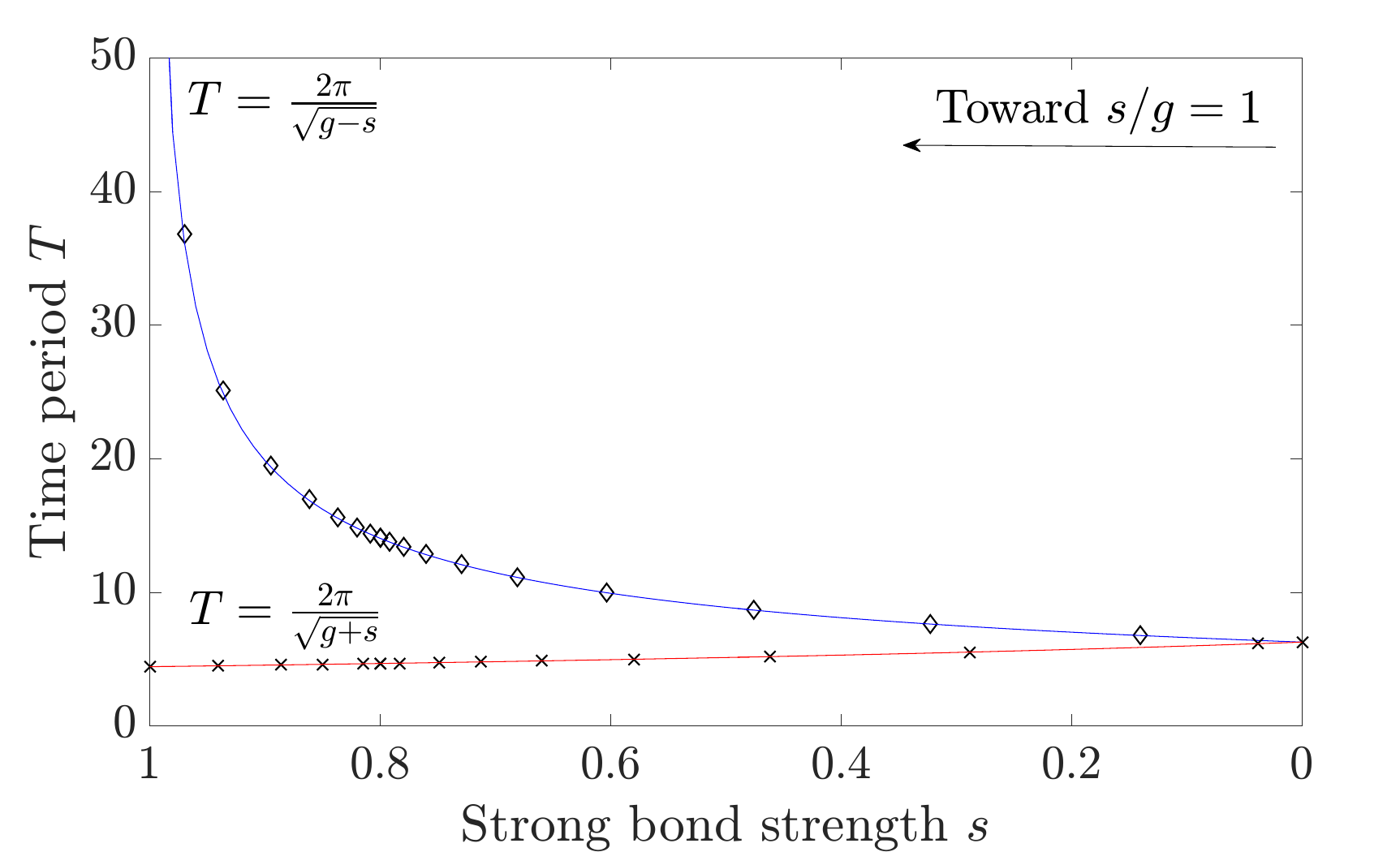}
    \caption{Comparison of time periods of fully dimerized sub-circuits and bulk nodes in the SSH chain. Solid curves denote theoretical time period for in-phase and anti-phase modes of the dimer. Diamonds and crosses denote time periods of bulk nodes in the SSH chain at different values of $s$.}
    \label{fig:S1}
\end{figure}

We see that the eigenmodes of the decoupled dimers and boundary nodes are consistent with the eigenmodes of the whole chain. In Fig. \ref{fig:S1}, there is a strong agreement between the time period of fully dimerized sub-circuits and that of the bulk nodes in the SSH chain. Thus, we have validation that taking the fully dimerized limit is in fact accurate and hence, can prove to be advantageous as shown. It also explains why in the topological regime, we have characteristically different behavior for the boundary (isolated nodes) and the bulk (dimers).

In the same way, we extend this theory for the 2D stacked SSH sample in which we analyze bulk quartets. The internal dynamics is given by a ring of van der Pol oscillators
\begin{subequations}
\begin{align}
	\ddot{x}_1 - \varepsilon (1 - x_1^2) \dot{x}_1 + g x_1 =  s x_2 + s x_4, \\
	\ddot{x}_2 - \varepsilon (1 - x_2^2) \dot{x}_2 + g x_2 =  s x_3 +  s x_1, \\
	\ddot{x}_3 - \varepsilon (1 - x_3^2) \dot{x}_3 + g x_3 =  s x_4 + s x_2, \\
	\ddot{x}_4 - \varepsilon (1 - x_4^2) \dot{x}_4 + g x_4 =  s x_1 + s x_3.
\end{align}
\end{subequations}
Once again, it can be shown that the eigenmodes of the coupling matrix include the in-phase state (denoted by $\begin{pmatrix} 1 & 1; & 1 & 1  \\ \end{pmatrix}$) and the anti-phase states (either $\begin{pmatrix} 1 & -1; & -1 & 1  \\ \end{pmatrix}$ or $\begin{pmatrix} 1 & 1; & -1 & -1  \\ \end{pmatrix}$ depending on which topological regime the network is in). It can be shown that the in-phase state is more stable than the anti-phase states, thereby favoring synchronization in the bulk. The natural frequencies of the bulk are calculated to be $\sqrt{g \pm s \pm s}/(2 \pi)$.

\section{Frequency control}

\begin{figure}[t]
    \centering
    \includegraphics[width=.9\textwidth]{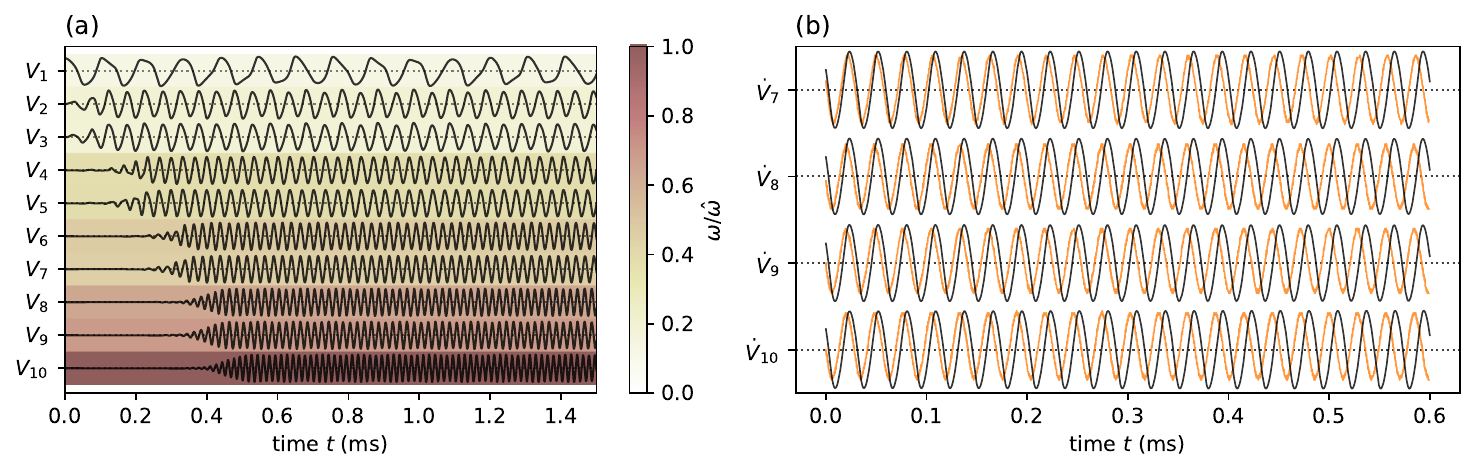}
     \caption{Frequency control in a 1D ATC. (a) Theoretical time series of 10-node 1D SSH chain with localized frequency pattern. Darkening of background color denotes frequency increase along length of the chain.
     (b) Experimental and theoretical time series comparison for the low-energy in-phase attractor of 2D SSH sample in T2-corner mode regime. Since all ground inductances are equal, natural frequencies of all nodes are equal resulting in synchrony.
     }
    \label{fig:freq-control}
\end{figure}

We generalize the coupling matrix in Eq.~(3) of the main text, by replacing $g$ by $g_i$ to get
\begin{align}
\boldsymbol{\beta} = \begin{pmatrix} g_1 & -w & 0 & 0 & 0 & 0 \\
 				-w & g_2 & -s & 0 & 0 & 0 \\
				 0 & -s & g_3 & -w & 0 & 0 \\
				 0 & 0 & -w & g_4 & -s & 0 \\
				 0 & 0 & 0 & -s & g_5 & -w  \\
				 0 & 0 & 0 & 0 & -w & g_6
				 \end{pmatrix},
\end{align}
where $g_i =  L_0(1/L_{ii} + \sum_{ \substack{ k \\ k \neq i } } 1/L_{ik})$ such that $L_{ii}$ can be chosen independently for each $i$. Since the frequencies of the dimers and quartets are proportional to $\sqrt{g_i \pm s}$ and $\sqrt{g_i \pm s \pm s}$ respectively, by varying $L_{ii}$ appropriately, we can precisely control the frequency of each of the isolated nodes, dimers and quartets. We exploit this technique to construct two ATCs with specific frequency patterns.

In Fig. \ref{fig:freq-control} (a), we construct a 1D SSH chain of 10 nodes where the ground inductances are chosen such that there is an increase in frequency by $10\, \mathrm{k Hz}$ with each successive dimer/isolated node along the length of the chain. We obtain a precise, localized frequency pattern as can be seen via the background color darkening along the length of the chain.

In Fig. \ref{fig:freq-control} (b), we consider a square 2D SSH sample in the T2-corner mode regime. Choosing the ground inductances of all nodes to be equal, say $L$, we get the same in-phase frequency for all of the isolated nodes, dimers and quartets. This can be seen via the following computations:
\begin{subequations}
\begin{eqnarray}
   g_c &=& L_0 \bigg( \frac{1}{L} + \bigg( \frac{2}{L_w} \bigg) \bigg), \quad\\
    g_e &=& L_0 \bigg( \frac{1}{L} + \bigg( \frac{2}{L_w} + \frac{1}{L_s} \bigg) \bigg), \quad\\
    g_b &=& L_0 \bigg( \frac{1}{L} + \bigg( \frac{2}{L_w} + \frac{2}{L_s} \bigg) \bigg), \quad\\
    s &=& \frac{L_0}{L_s},\\
    g_c &=& g_e - s = g_b - 2s = L_0 \bigg( \frac{1}{L} + \frac{2}{L_w} \bigg),
\end{eqnarray}
\end{subequations}
where the subscripts $c$, $e$ and $b$ stand for corner (isolated), edge (dimer) and bulk (quartet) respectively. This leads to a probabilistically favored synchronized state where the frequency of all nodes are equal.
For the experiment, we make use of the same parameters used in Tab. \ref{tab:components} except that we include ground inductances for the bulk nodes this time. Once again, due to its topological nature, such states are robust against lattice defects.

\section{Topological edge and corner modes}

\begin{figure*}[h]
    \centering
    \includegraphics[width=\textwidth]{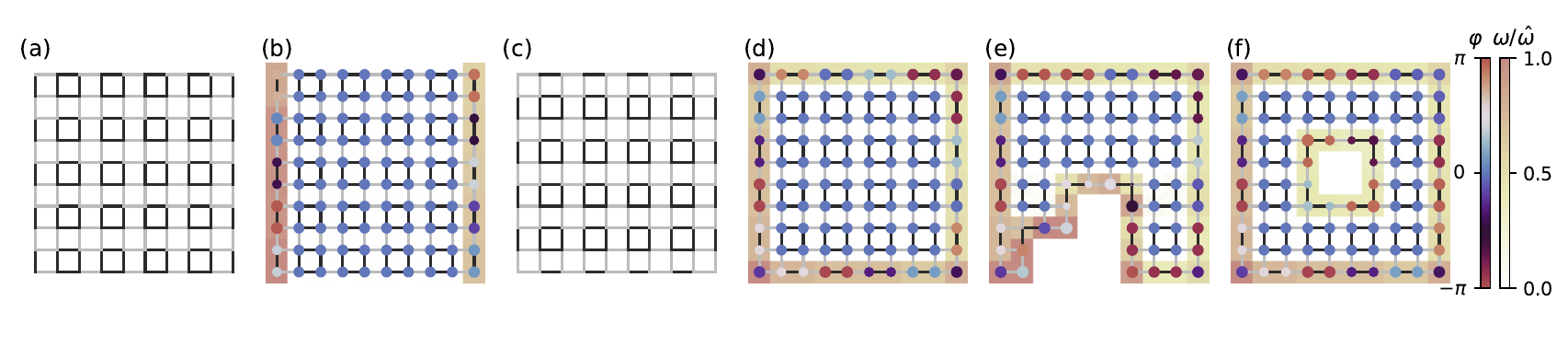}
    \caption{Self-organized self-sustained nonlinear oscillations  in 2D ATCs recapitulate
    topological boundary phenomena.
    (a)~Connectivity of the 2D SSH grid in the
    topological T1-edge mode regime. Thick and thin lines indicate strong and weak couplings, respectively.
    (b)~Snapshot of the vdP network dynamics on the grid (a), with nodes colored by the phase angle $\varphi$ in the $(x_i,\dot x_i)$ plane; background color shows time averaged node frequency
    $\omega=\langle\dot\varphi\rangle$ normalized by the maximum $\hat\omega$ over all oscillators.
    Even though the initial voltage was non-zero only on the bottom left corner node, the circuit autonomously actuates two topological edge oscillations with mean frequency $\hat \omega$, while the bulk synchronizes in-phase.
    (c)~Connectivity of the 2D SSH grid in the
    topological T2-corner mode regime.
    (d)~Snapshot of the vdP network dynamics on the grid (c), for the same corner-localized  initial conditions as in (b). This circuit self-organizes into a stable active state in which edges and corners oscillate while the bulk synchronizes in-phase.
    (e,f)~Snapshot of the vdP network dynamics on the grid (c) with edge (e) and bulk (f) defects. The vdP oscillator network remains polarized in the bulk, while active topological modes wrap around the defect. Simulation parameters: $\varepsilon=0.1, g=1,  w=5 \times 10^{-6}, s=g/2-w$. Initial conditions: $x_1(0)=2$ and $\dot{x}_1(0)=x_j(0)=\dot{x}_j(0)=0$ for $j\ge 2$.}
    \label{fig:edges_corners}
\end{figure*}

In Fig.~\ref{fig:edges_corners} we show the results of numerical
simulations of the two major configurations of 2D SSH models,
T1 and T2. The major features of passive topological excitations
are preserved in the nonlinear, active regime.

\section{Oscillatory pattern control}

\begin{figure}[h]
    \centering
    \includegraphics[width=0.4\columnwidth]{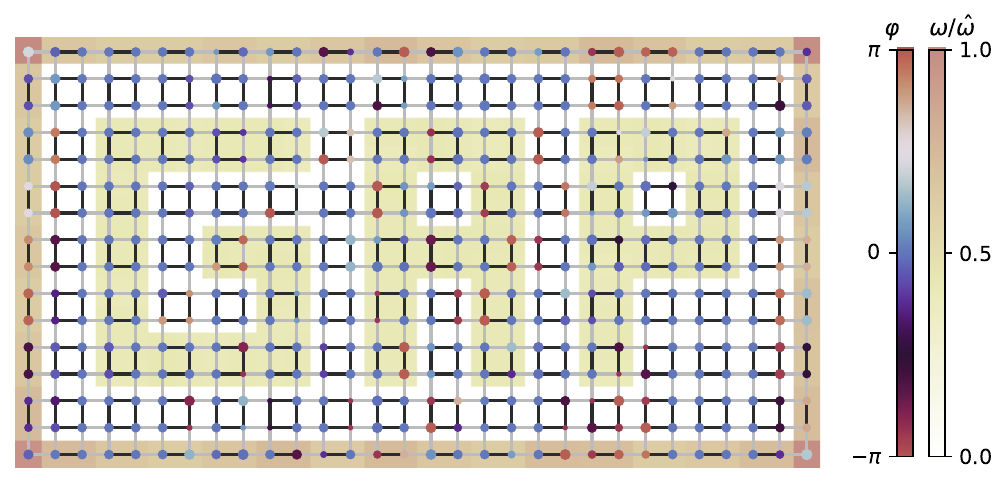}
    \caption{Control of active traveling wave patterns through bulk defects. Activated in a single corner, the circuit self-organizes into a pre-programmed wave pattern (SI Movie 5).  Simulation parameters: $\varepsilon=0.5, g=1, w=5 \times 10^{-6}, s=g/2-w$. Initial conditions: $x_1(0)=2$ and $\dot{x}_1(0)=x_j(0)=\dot{x}_j(0)=0$ for $j\ge 2$.}
    \label{fig:defects}
\end{figure}

By selectively switching strong to
weak bonds in the bulk, it is possible to introduce local
defects which self-excite into higher frequency active oscillations
that can be used to design large-scale patterns (Fig.~\ref{fig:defects}).

\section{Activity only on the boundary}
\begin{figure}[h]
    \centering
    \includegraphics[width=\textwidth]{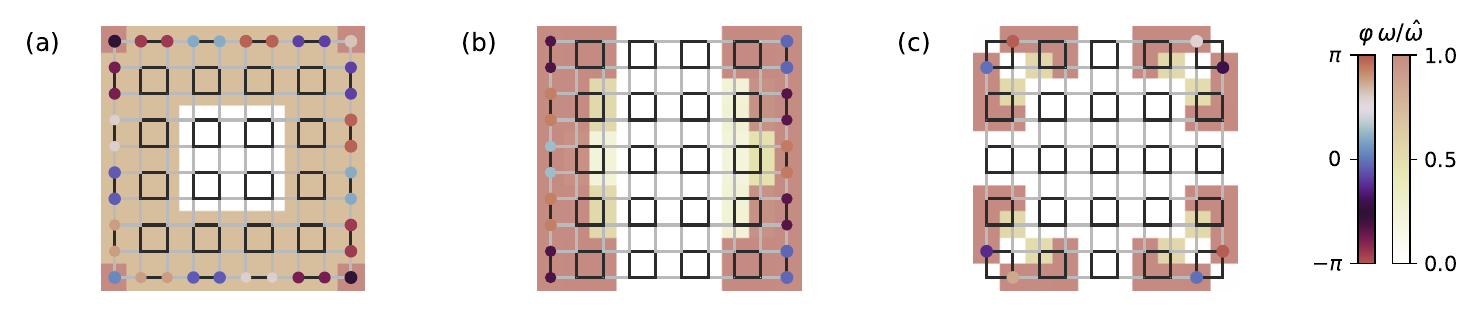}
    \caption{Global topological states on 2D ATCs with activity only on the boundary (cf. Fig.~4).
    (a)~Snapshot of the vdP network dynamics in the T2-corner mode regime, with nodes colored by the phase angle $\varphi$ in the $(x_i,\dot x_i)$ plane; background color shows time averaged node frequency
    $\omega=\langle\dot\varphi\rangle$ normalized by the maximum $\hat\omega$ over all oscillators.
    Even though the initial voltage was non-zero only on one node in the bulk, the circuit autonomously actuates the entire boundary with exponential decay in amplitude along the bulk.
    (b)~Snapshot of the vdP network dynamics in the T1-edge mode regime. The circuit autonomously actuates two topological edge oscillations with exponential decay in amplitude along the bulk.
    (d)~Snapshot of the vdP network dynamics in the non-topological regime. The circuit autonomously actuates nodes adjacent to the corners, but this state is neither topological nor global.}
    \label{fig:only-boundary}
\end{figure}

Global patterns can be realized in classical topological insulators by adding activity only to the boundary. This results in patching up of the localized topological modes giving rise to a global, robust state. Moreover, these states can be excited from the bulk as seen in the following simulations. We investigate by considering a $10 \times 10$ 2D SSH sample in the T1-edge mode regime, T2-corner mode regime and the non-topological regime (Fig. \ref{fig:only-boundary}). Upon enabling activity only on the boundary, we excite the global topological state via an initial condition in the bulk. We observe that the samples in the topological regimes (a,b) display the same topological patterns with the added feature of exponential decay in amplitude as seen in classical passive systems. However, for the non-topological regime (c), we obtain oscillations only on nodes adjacent to the corners, indicating that global patterns in ATCs are not a consequence of adding activity alone; topology is also essential.

\begin{figure}
    \centering
    \includegraphics[width=.6\textwidth]{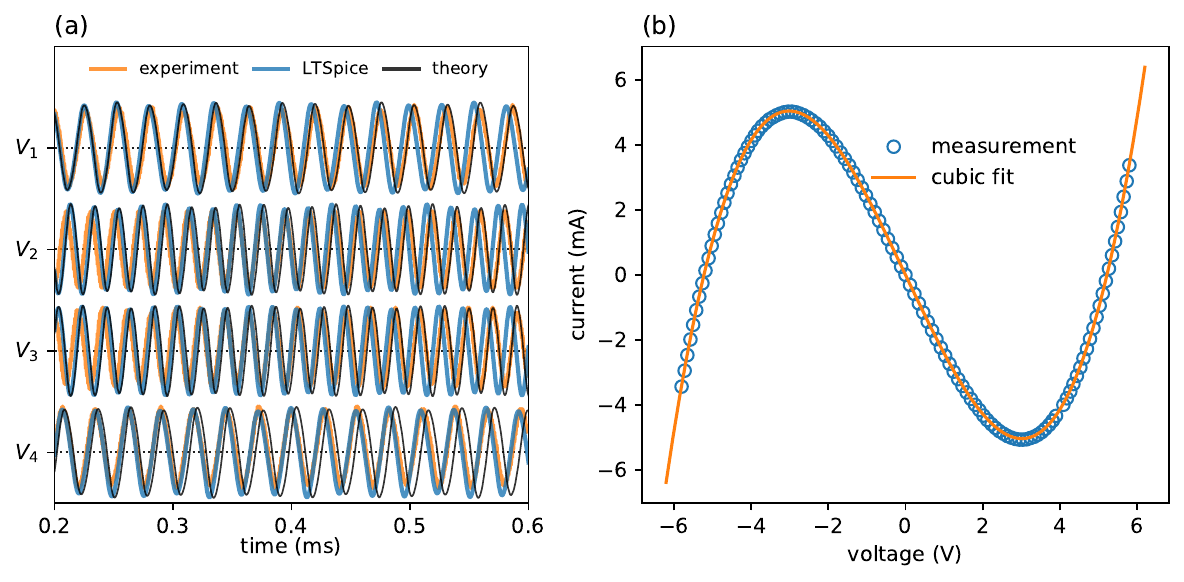}
    \caption{
    (a) Experimental and theoretical time series comparison for the high-energy anti-phase attractor. The initial conditions on
    $V_2$ and $V_3$ had opposite signs.
    Orange lines correspond to experimental data, blue lines to LTSpice simulations and
    black lines to numerical simulations using Eq.~(2) from the main paper.
    All voltages vary between $\pm 6\, \mathrm{V}$.
    (b) Measured $I$-$V$ characteristic curve of the Chua circuit.
    We applied voltages between $-6\,\mathrm{V}$ and $+6\,\mathrm{V}$
    to the circuit and measured the current through the element, obtaining
    a cubic relationship.
    The fitted curve corresponds to the model $I(V) = -2.5\,\mathrm{mS}\; V+ 0.091\,
    \mathrm{mS/V}^{2}\; V^3$.}
    \label{fig:si-anti}
\end{figure}

\section{Experimental realization using Chua's circuit}
Our first experimental realization of an active topolectrical circuit implements
a 4-node version of the circuit in Fig.~1(b) for nodes $V_1$ to $V_4$.
The individual circuit elements are off-the-shelf components from Digi-Key Electronics (Digi-Key Electronics, Thief River Falls, MN, U.S.A).
For the active nonlinear resistance, we constructed a realization of
Chua's diode~\cite{zhong1994} [Fig.~2(a)], which features a cubic current-voltage relationship.
We measured the current through the Chua circuit when applying a voltage
$V_i$ and found almost perfectly cubic behavior [Fig.~\ref{fig:si-anti}(b)].
We fitted the curve to a cubic polynomial and discarded the small
constant and quadratic terms, which resulted in the simplified model
$I(V) = -\alpha V + \gamma V^3$, which is also shown in Fig.~\ref{fig:si-anti}(b).
The parameters of all circuit elements are collected in
Tables~\ref{tab:param} and \ref{tab:components}.
\begin{table}[]
\centering
\begin{tabular}{|c|c|c|c|}
\hline
\textbf{Dimensional parameter} & \textbf{Value}             & \textbf{Nondimensional parameter} & \textbf{Value} \\ \hline
$C$                            & $42\, \mathrm{nF}$           & $g$                               & $1.05$         \\ \hline
$L_e$                          & $525\, \mathrm{\mu H}$       & $w$                               & $0.05$         \\ \hline
$L_b$                          & Nil                        & $s$                               & $1.0$            \\ \hline
$L_w$                          & $10.5\, \mathrm{mH}$         & $\varepsilon$                     & $0.28$         \\ \hline
$L_s$                          & $525\, \mathrm{\mu H}$       &                                   &                \\ \hline
$\alpha$                       & $2.5\, \mathrm{m S}$         & \textbf{Scaling factor}                                  &    \textbf{Value}            \\ \hline
$\gamma$                       & $0.091\, \mathrm{mS/V^{2}}$ & $V_0$                                  & $3.03\, \mathrm{V}$               \\ \hline
$R_1$                          & $2\, \mathrm{k \Omega}$    & $\tau$                                  & $4.7\, \mathrm{\mu s}$               \\ \hline
$R_2$                          & $2\, \mathrm{k \Omega}$    & $L_0$                                  &    $525\, \mathrm{\mu H}$            \\ \hline
$R_3$                          & $400\, \mathrm{ \Omega}$    &                                   &                \\ \hline
$R_4$                          & $3010\, \mathrm{ \Omega}$   &                                   &                \\ \hline
$R_5$                          & $7900\, \mathrm{ \Omega}$   &                                   &                \\ \hline
\end{tabular}
\caption{Realistic parameter values for the topological regime for a 1D SSH chain. $L_w$ and $L_s$ denote the inductances in the weak and strong couplings respectively while $L_e$ and $L_b$ denote the ground inductances of the edge and bulk nodes. Note that for the bulk nodes, we do not use ground inductances.}
\label{tab:param}
\end{table}

\begin{table}[]
\centering
\begin{tabular}{|c|c|c|c|c|}
\hline
\textbf{Component/Parameter} & \textbf{Node 1}              & \textbf{Node 2}              & \textbf{Node 3}              & \textbf{Node 4}                     \\ \hline
$\alpha$                     & $2.496\, \mathrm{mS}$          & $2.494\, \mathrm{mS}$          & $2.500\, \mathrm{mS}$          & $2.500\, \mathrm{mS}$          \\ \hline
$\gamma$                     & $0.09094\, \mathrm{mS/V^{2}}$ & $0.09112\, \mathrm{mS/V^{2}}$ & $0.09200\, \mathrm{mS/V^{2}}$ & $0.09092\, \mathrm{mS/V^{2}}$ \\ \hline
$L_{ii}$                       & $485.6\, \mathrm{\mu H}$       & Nil                          & Nil                          & $486.0\, \mathrm{\mu H}$        \\ \hline
$L_{ii}$ ESR                 & $0.821\, \mathrm{\Omega}$      & Nil                          & Nil                          & $0.810\, \mathrm{\Omega}$          \\ \hline
$C$                          & $41.99\, \mathrm{n F}$         & $41.63\, \mathrm{n F}$         & $42.07\, \mathrm{n F}$         & $42.06\, \mathrm{n F}$         \\ \hline
$C$ ESR                      & $0.459\, \mathrm{\Omega}$      & $0.620\, \mathrm{\Omega}$      & $0.704\, \mathrm{\Omega}$     & $0.760\, \mathrm{\Omega}$      \\ \hline
$L_w$                        & \multicolumn{2}{c|}{$10.5020\, \mathrm{m H}$}                 & \multicolumn{2}{c|}{$10.5060\, \mathrm{m H}$}                     \\ \hline
$L_w$ ESR                    & \multicolumn{2}{c|}{$2.520\, \mathrm{\Omega}$}                & \multicolumn{2}{c|}{$2.528\, \mathrm{\Omega}$}                   \\ \hline
$L_s$                        &                              & \multicolumn{2}{c|}{$512.0\, \mathrm{\mu H}$}                 &                                    \\ \hline
$L_s$ ESR                    &                              & \multicolumn{2}{c|}{$0.940\, \mathrm{\Omega}$}                &                                     \\ \hline
\end{tabular}
\caption{Component values used for experiments in Figs.~2 and \ref{fig:si-anti}. ESR stands for equivalent series resistance, referring to the parasitic resistance present in practical, non-ideal inductors and capacitors. For the values listed, ESR was measured at an AC frequency of 1 kHz.}
\label{tab:components}
\end{table}

The circuits were powered using two Extech 382260 80W switching DC power
supplies (FLIR Commerical Systems, Nashuah, NH, U.S.A.), and
measurements were taken using a Rigol DS 1054Z 4-channel digital
oscilloscope (Rigol Technologies U.S.A., Beaverton, OR, U.S.A.).

Fig.~\ref{fig:si-anti}~(a) shows an alternate dynamical attractor which is the high-energy bulk oscillation obtained using anti-phase initial conditions. This state is achieved using the same experimental setup in Fig.~2.

\section{Alternative experimental realization}

\begin{figure}[h]
\centering
 \includegraphics[width=\textwidth]{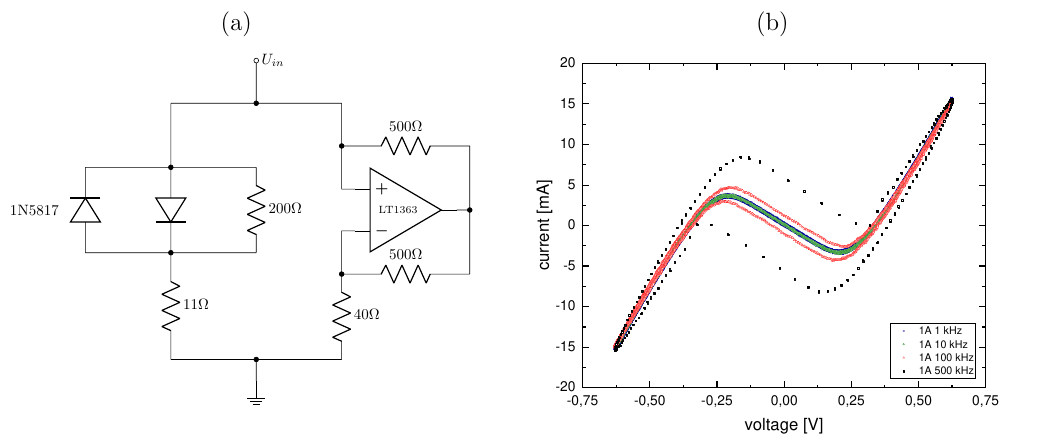}
 \caption{Schematic and characterization of the Van der Pol resistor element. (a) Circuit diagram of the Van der Pol nonlinear resistor, using two parallel diodes in opposite orientation and a negative impedance converter. (b) Current-voltage characteristic of the Van der Pol element used in node 1A of the 1D-chain, measured for different AC voltage frequencies. For large frequencies $>100~\si{\kilo\hertz}$, significant hysteresis is observed.}
\label{Fig:Schematic}
\end{figure}

Our alternative implementation of the Van der Pol SSH circuit is based on the nonlinear resistor suggested in  Ref.~\cite{Makarov_2001}. Fig.~\ref{Fig:Schematic}~(a) shows a circuit diagram of the resistor. The negative impedance converter provides the negative resistance at low voltages, corresponding to the negative linear term, while the two diodes approximate the nonlinear contribution.
Fig.~\ref{Fig:Schematic}~(b) shows the circuit's current-voltage characteristic, measured by applying an AC voltage for different frequencies. A third order polynomial fit $I(V)=-\alpha V + \gamma V^3$ to the $10\,\mathrm{kHz}$ curve in the range of $-0.3$ V to $0.3$ V gives parameter values $\alpha=(0.02494 \pm 0.00007)\, \mathrm{\Omega^{-1}}$ and $\gamma = (0.2069\pm0.00010) \mathrm{\Omega^{-1} V^{-2}}$.
At large frequencies $ >100~\si{\kilo\hertz}$, significant hysteresis occurs. In preliminary simulations of a 1D oscillator chain, this behavior has shown to cause problems at frequencies beyond $1\,\si{\mega\hertz}$ which has to be taken into consideration in the choice of the circuit's remaining parameters.

\begin{figure}[tp]
	\includegraphics[width=.9\textwidth]{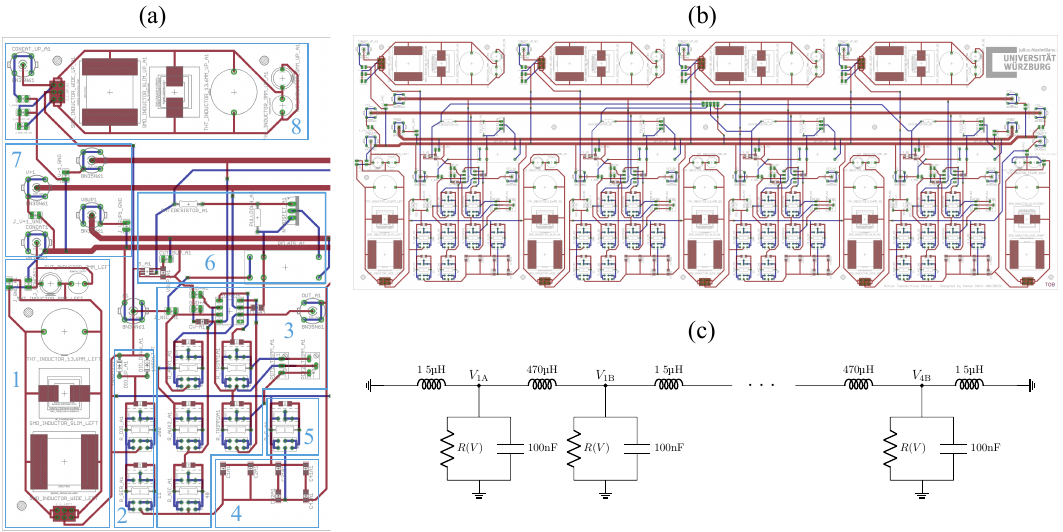}
	\caption{(a) Illustration of the circuitry of the first node. 1: Footprints to solder different kinds of inductors, 2: anti-parallel diodes with serial and parallel resistor to implement nonlinearity, 3: OpAmp circuit used as negative impedance converter, 4: grounding capacitors, 5: grounding resistor (not used), 6: two kinds of switches to connect the node to ground automatically, analog switch integrated circuit (IC) (left), MOSFET driven analog relay (right), 7: BNC connectors to link several boards together, (from top to bottom) negative and positive power supply for the INICs, power supply for the switches and ground line, 8: additional spot for inductors to expand the circuit into the second dimension. (b) Illustration of an entire circuit board module. It hosts 4 nodes with the ability to expand the network into the second dimension. (c): Circuit diagram of the 1D vdP-SSH chain. R(V) represents the nonlinear resistor element.}
	\label{Fig:CircuitBoard}
\end{figure}

We designed modular circuit boards, each containing four nodes, that can be connected to form 1D and 2D lattices. Fig.~\ref{Fig:CircuitBoard}~(a) shows an illustration of a section of the board, that corresponds to one node, Fig.~\ref{Fig:CircuitBoard}~(b) an entire four-node module. For our measurements, we connect these boards to an 8-site 1D chain and an 8x8 sites 2D lattice. In both setups, the strong coupling is represented by inductors $L_s=1.5\,\si{\micro\henry}$ (Murata 15152C) and the weak couplings by inductors $L_w=470\,\si{\micro\henry}$ (Bourns RL187-471J-RC). In the 1D circuit, all nodes are grounded by $100\,\si{\nano\farad}$ capacitors (Murata GCM31C5C1H104FA16L), in the 2D circuit by $39\,\si{\nano\farad}$ capacitors (Vishay VJ1206A393FXJTW1BC). This choice of circuit elements correspond to the non-dimensional parameter values of $g=1$, $w=0.0032$, $s=g-w=0.9968$ and $\epsilon=0.0964$ in the 1D setup. The corresponding values in the 2D setup are $g=1$, $w=0.0016$, $s=(g-w)/2=0.4984$ and $\epsilon=0.1092$.

\begin{figure}[p]
\centering
 \includegraphics[width=.9\textwidth]{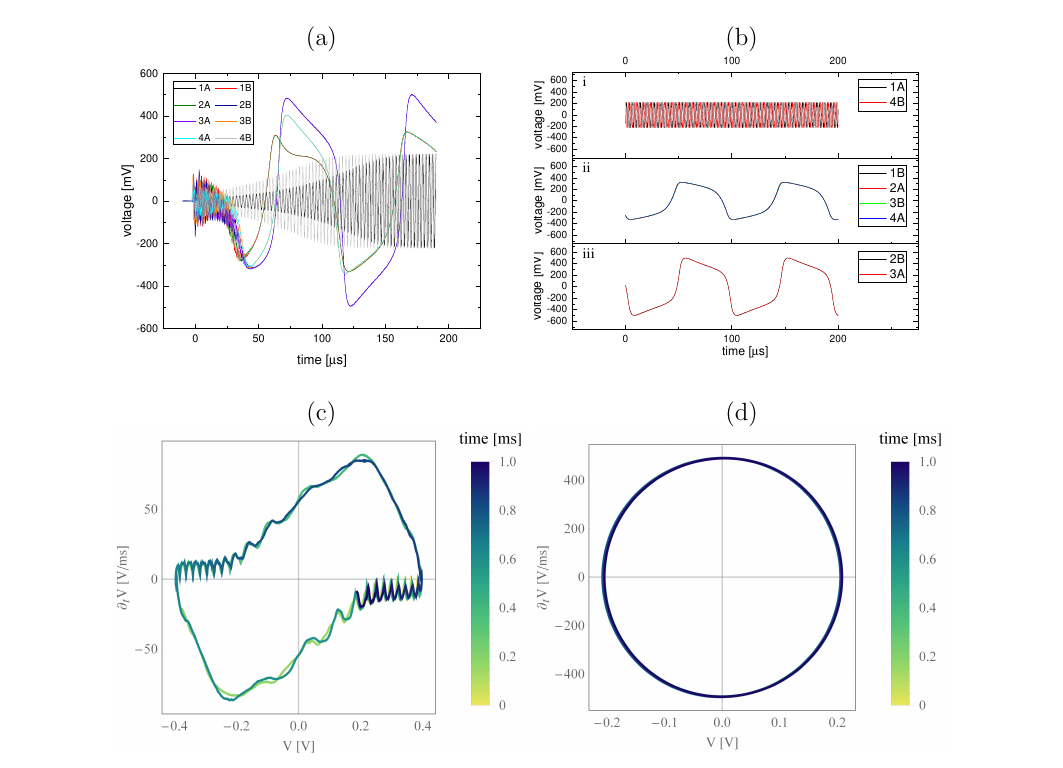}
\caption{Measurement results of the 1D VdP-SSH chain. (a): Transient voltage signals after the operational amplifiers' supply voltage is connected. (b): Steady-state oscillatory patterns of the circuit. The two boundary nodes 1A and 4B oscillate at a relatively high frequency. The bulk mode voltages oscillate at considerably lower frequency and are notably distorted. The two tightly bound pairs adjacent to the boundaries, (1B, 2A) and (3B, 4A) show almost identical voltages, while the two central nodes (2B, 3A) show slightly larger amplitudes. (c) and (d): Phase space diagrams $V$, $\dot{V}$ of bulk node 3A (c) and boundary node 1A(d). Bulk node 3A shows a strongly nonlinear limit cycle, while the trajectory of boundary node 1A does not visibly deviate from that of a linear oscillator.}
\label{Fig:meas_1D}
\end{figure}

Circuit diagram Fig.~\ref{Fig:CircuitBoard}~(c) shows the 1D circuit's layout. The nodes in the 1D circuit are labeled 1A, 1B, 2A, ... 4B, with the number indicating the unit cell number and the character the sublattice site.
When the active elements' supply voltage is connected, the circuit is excited by background noise from the environment or active components and quickly self-organizes to a steady oscillatory pattern, see Fig.~\ref{Fig:meas_1D}~(a). Fig.~\ref{Fig:meas_1D}~(b) shows the different resulting oscillation patterns. As expected, the boundary sites oscillate at a relatively high frequency (approx. 380 kHz) compared to the bulk sites. The tightly coupled sites 1B and 2A, 2B and 3A, ect. show nearly identical voltage signals. There is an observable difference for the central pair of nodes 2B and 3A, that exhibit a larger oscillation amplitude than the neighboring pairs. Comparing the phase space diagrams of a bulk node (3A, Fig.~\ref{Fig:meas_1D}~(c)) and a boundary node (1A, Fig.~\ref{Fig:meas_1D}~(d)) confirms that the former follows a strongly nonlinear limit cycle, while the latter oscillates nearly harmonically.

\begin{figure}[p]
    \centering
    \includegraphics[width=.9\textwidth]{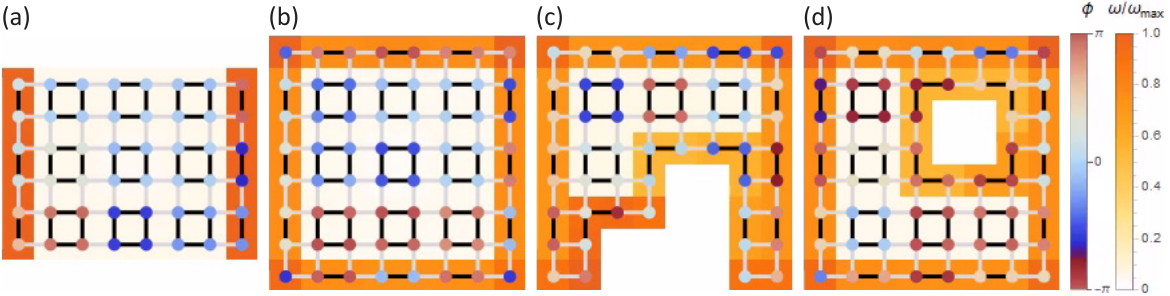}
    \caption{Synchronized time-resolved measurements of the $8 \times 8$ 2D active topolectrical lattice in different topological regimes. The color of the nodes (circles) displays the phase angle between the normalized voltage and its derivation. The background colors indicate the normalized time average of the node's angular frequency. (a)~In the T1-edge mode regime the circuit shows synchronized oscillations of the strongly coupled edge dimers with high frequency at the topological edges. Each bulk quartet with strong coupling oscillates phase synchronized but with significantly lower frequency than the edge dimers. Due to the components tolerances there are small differences in the oscillation frequencies between the individual bulk quartets preventing the synchronization of all bulk nodes. (b)~In the T2-corner mode regime all edges host fast oscillating edge dimers while the corners of the lattice are connected to their neighbouring nodes only by weak couplings resulting in corner oscillators with oscillation frequencies higher than the edge dimers. (c,d)~Defects at the edge (c) or in the bulk (d) do not amend the self-organization of the active lattice.}
    \label{Fig:alternative_2D_phases}
\end{figure}

Next, we examine the 2-dimensional setup.
To investigate the self-organized self-sustained nonlinear oscillations shown in the simulation in Fig.~\ref{fig:edges_corners} a time-resolved measurement of all lattice nodes in different topological regimes is shown in Fig.~\ref{Fig:alternative_2D_phases}. The voltages were measured by 9 synchronized 8-channel PC hosted oscilloscopes (PicoScope 4824) and numerically differentiated to calculate the phase angle $\varphi$ in the $(x_i,\dot x_i)$ plane and the time averaged node frequency $\omega=\langle\dot\varphi\rangle$. The lattice is investigated in two topological regimes. In the T1-edge mode regime Fig.~\ref{Fig:alternative_2D_phases}~(a) two edges (left and right) of the 2D lattice are terminated to host topological edge modes. In agreement with the simulation results in Fig.~\ref{fig:edges_corners}~(b) the topological edges host strongly coupled edge dimers which show synchronized high frequency oscillation (in respect to the bulk). In the bulk each four nodes connected by strong coupling form synchronized quartets oscillating with low frequency. The edge oscillator's limit cycles are approximately circular whereas the bulk limit cycles are of strong relaxation type.

Contrary to the simulation results the bulk quartets are not completely in-phase due to slight variations of the oscillation frequencies induced by the components tolerances. Beside the slowly oscillating bulk quartets and the rapidly oscillating edge dimers described before, the T2-corner mode lattice Fig.~\ref{Fig:alternative_2D_phases}~(b) hosts corner modes at it's only weakly coupled corners with approximately circular limit cycles and oscillation frequencies even higher then those of the edge dimers. As it can be seen in Fig.~\ref{Fig:alternative_2D_phases}~(c,d) neither an edge nor a bulk defect can affect the ATC self-organization principle. In accordance to the simulations Fig.~\ref{fig:edges_corners}~(e,f) the bulk stays organized in synchronized bulk quartets while the boundary mode wraps around the defect and keeps oscillating with high frequency.

\end{document}